\documentclass[journal,draftclsnofoot,onecolumn, 12pt]{IEEEtran}

\usepackage[T1]{fontenc}

\usepackage{cite}
\ifCLASSINFOpdf
   \usepackage[pdftex]{graphicx}
   \DeclareGraphicsExtensions{.pdf}
\else
   \usepackage[dvips]{graphicx}
   \DeclareGraphicsExtensions{.eps}
\fi
\usepackage[cmex10]{amsmath}
\usepackage{amssymb}
\usepackage{mathtools}

{}
\newtheorem{lemma}{Lemma}
\newtheorem{proposition}{Proposition}

\usepackage[font=footnotesize, caption=false]{subfig}
\begin{document}
%
% paper title
% can use linebreaks \\ within to get better formatting as desired
\title{Asymptotic Analysis of Multi-Antenna Cognitive Radio Systems Using Extreme Value Theory}
%
%
% author names and IEEE memberships
% note positions of commas and nonbreaking spaces ( ~ ) LaTeX will not break
% a structure at a ~ so this keeps an author's name from being broken across
% two lines.
% use \thanks{} to gain access to the first footnote area
% a separate \thanks must be used for each paragraph as LaTeX2e's \thanks
% was not built to handle multiple paragraphs
%

\author{Ruifeng~Duan,~~\IEEEmembership{Member,~IEEE,}
				Zhong~Zheng,~\IEEEmembership{Member,~IEEE,}
        Riku~J\"{a}ntti,~\IEEEmembership{Senior Member,~IEEE,}
        %and~ Mohammed~S.~Elmusrati,~\IEEEmembership{Senior Member,~IEEE,}
        and~Jyri~H\"{a}m\"{a}l\"{a}inen,~\IEEEmembership{Member,~IEEE}% <-this % stops a space
%\thanks{Manuscript submitted \today} %; revised \today, 2012.}
\thanks{The authors are with the Department of Communications and Networking, Aalto University, Finland. (emails: ruifeng.duan, zhong.zheng, riku.jantti, jyri.hamalainen@aalto.fi).}
%\thanks{This work was supported in part by $\cdots$.}
}

\maketitle

\begin{abstract}
%\boldmath
We consider a spectrum-sharing cognitive radio system with antenna selection applied at the secondary transmitter (ST). Based on the extreme value theory, we deduce a simple and accurate expression for the asymptotic distribution of the signal to interference plus noise ratio at the secondary receiver. Using this result, the asymptotic mean capacity and the outage capacity for the secondary user (SU) are derived. The obtained asymptotic capacities approach the exact results as the number of transmit antennas $N$ increases. Results indicate that the rate of the SU scales as $\log(N)$ when the transmit power of the ST is limited by the maximum allowable interference level, while rate scales as $\log(\log(N))$ if ST is limited by the maximum transmit power.

\end{abstract}
% IEEEtran.cls defaults to using nonbold math in the Abstract.
% This preserves the distinction between vectors and scalars. However,
% if the journal you are submitting to favors bold math in the abstract,
% then you can use LaTeX's standard command \boldmath at the very start
% of the abstract to achieve this. Many IEEE journals frown on math
% in the abstract anyway.

% Note that keywords are not normally used for peerreview papers.

% For peer review papers, you can put extra information on the cover
% page as needed:
 \ifCLASSOPTIONpeerreview
 \begin{center} \bfseries EDICS Category: 3-BBND \end{center}
 \fi
%
% For peerreview papers, this IEEEtran command inserts a page break and
% creates the second title. It will be ignored for other modes.
\IEEEpeerreviewmaketitle

% %
\section{Introduction}

In order to utilize the radio spectrum more effectively, the secondary users (SUs) may be allowed to share the spectrum with the primary users (PUs). However, the SUs need to control the transmit power in order not to cause harmful interference to the PUs \cite{Goldsmith2009}. While research literature on this topic is rich, we recall in the following works that are most relevant for this study. Results in \cite{Gastpar2007} have shown that, under a received-power constraint, the channel capacity of the SU can be significantly different from the one in a conventional AWGN channel. In \cite{Kang2009}, authors studied the ergodic capacity and the outage capacity of the SU assuming fading channels, where the interference from the PU to the SU was ignored. In \cite{Suraweera2010}, authors studied a single-antenna spectrum-sharing system by taking into consideration the interference from the PU at the secondary receiver (SR). 

The multiple antenna techniques provide an efficient tool to improve the link level capacity. Among others, the transmit antenna selection (TAS) scheme is attractive, since it can reduce the hardware costs and complexity \cite{Sanayei2004}. In \cite{Blagojevic2012}, authors studied the ergodic capacity of the SU using TAS and maximum ratio combining (MRC) at the receiver. However, the PU interference was ignored, and no closed-form expression was given for the SU ergodic capacity in the presence of the peak transmit power and interference power constraints. Authors in \cite{Tourki2014} derived the exact expressions for the ergodic capacity of MIMO cognitive system in terms of Fox's H-function. This result, although of theoretical interests, involves complicated contour integral and is thus difficult to use in practice.

In what follows, we consider a spectrum-sharing system with multiple antennas at the secondary transmitter. By assuming the interference from the PU to the SU and the transmit power constraints at the ST, we derive the asymptotic mean capacity and outage capacity for the link between ST and SR by applying the extreme value theory (EVT) \cite{Galambos1978}. The derived results are simple but accurate, and they become exact when the number of antennas approaches to infinity. Using these results, we obtain the rate scaling laws for the spectrum-sharing systems under the interference power and the transmit power limitations. 
%We observe that have the following contributions: 
%\begin{itemize}
	%%\item We proved that the signal to interference plus noise ratio of the secondary user lies on the maximum domain of the attraction of the Gumbel distribution.
	%\item The SU mean capacity and outage capacity have the same scaling characteristics following two scaling laws with respect to the number of antennas $N$. 
	%%First, it scales as $\log(N)$ in the interference power limited regime, i.e., $P_{\max}\rightarrow \infty$. Second, it scales as $\log(\log(N))$ in the transmit power limited regime, i.e., $P_{\max}$ is much smaller than the peak interference power threshold.
%\end{itemize}

The remainder of this paper is organized as follows. In Section \ref{sec:model}, we provide the system model and introduce the probability density function (PDF) and cumulative distribution function (CDF) of the SINR at the secondary receiver. In Section \ref{sec:EVT}, we briefly review the EVT, and show that the SU SINR of the proposed system lies on the domain of the attraction of the Gumbel distribution. Then we investigate the SU mean capacity, the SU outage capacity, and the scaling issues in Section \ref{sec:perf}. Section \ref{sec:Sim} presents the simulation results. The last Section concludes this paper.

%It is not straightforward to study the scaling behavior of the proposed scenario through providing analytical expressions. In this paper, we apply the extreme value theory (EVT), which has been applied into data analysis for extreme cases \cite{Galambos1978}. Recently, EVT has been adopted as a useful and powerful means for investigating the performance of the wireless communication networks. The authors in \cite{Song2006} conducted the asymptotic throughput analysis for channel-aware scheduling in an multiple-user scenario. The asymptotic antenna selection gain and the capacity distribution for a multiple antenna system were studied in \cite{Bai2009}. Heterogeneous feedback design issues were handled in \cite{Huang2013a}. For single-antenna over-laid cognitive radio (CR) systems, \cite{Hong2011} studied the multiuser diversity gain by applying EVT, where the detection of the PU transmission and the maximum transmit power constraint of the SU were taken into consideration. \cite{Xia2014} studied the limiting distribution function, diversity gain, and the coding gain of a single-antenna spectrum-sharing multi-hop relay system.

 %  \cite{Song2006, Bai2009, Hong2011, Huang2013a,Xia2014}
 %There may have, to the best of our knowledge, some unknown issues related the spectrum-shared systems that how the TAS technique improves the SU's performance when the primary user interference and the noise are considered, for instance, the SU mean capacity, the SU outage capacity, and the their scaling properties with respect to the number of antennas.

% %
\section{System and Channel Models}\label{sec:model}

\begin{figure}[!t]
\centering
\includegraphics[width=.4\columnwidth]{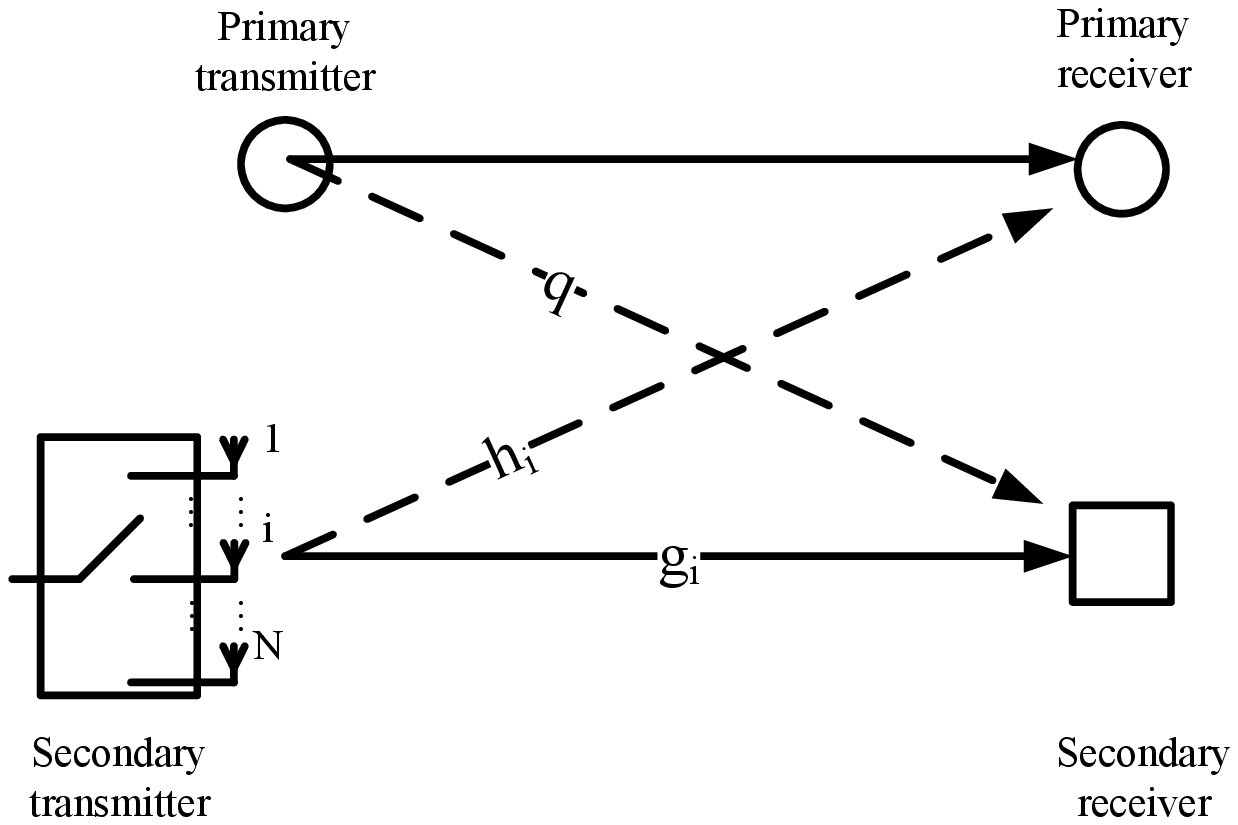}
\caption{Spectrum-sharing system with antenna selection at the secondary transmitter. Single antenna is equipped at the secondary receiver, the primary transmitter and receiver.}
\label{fig_SysMod}
\end{figure}

We consider a spectrum-sharing cognitive radio system depicted in Fig. \ref{fig_SysMod}. This model is similar to the one in \cite{Suraweera2010}, where the ST was equipped with single antenna. In our model, the secondary receiver (SR) is equipped with one antenna, and the secondary transmitter (ST) is equipped with $N$ antennas assuming TAS technique adopted at the ST. The primary transmitter (PT) and receiver (PR) are assumed to be equipped with one antenna. Let $g_i$, $h_i$ ($i = 1, \cdots, N$), and $q$ denote the channel power gains between the $i$-th antenna of the ST and the SR, the $i$-th antenna of the ST and the PR, and the PT and the SR, respectively. 

% Figure System Model

When the $i$-th antenna is used, the signal received at the SR, can be written as
\begin{equation}
	y_i =  \sqrt{P_s\, g_i}\, s + \sqrt{P_p\, q}\, z + \nu, \label{eq:signal_model}
\end{equation}
where $P_s$ and $P_p$ are the transmit power of the SU and PU, respectively, $s$ and $z$ denotes the transmit signals of the SU and PU with unit average power, and $\nu$ represents the additive white Gaussian noise with power $\sigma^2$. We assume that the channels experience independent block Rayleigh fading. Namely, the channels keep constant during each block, and are uncorrelated between blocks \cite{Ozarow1994}. Furthermore, the channels between the ST and the SR are identical and independent distributed (i.i.d.), and the channel between the ST and the PR is independent from the secondary channels. For a Rayleigh fading channel, the channel power gain $x$, $x\in \left\{g_i, h_i, q\right\}$, follows exponential distribution with mean value $\bar{x}$. The probability density function (PDF) of $x$ is given by
\begin{equation}
	f(x) = \frac{1}{\overline{x}}e^{-x/\overline{x}} , \quad x \geq 0
	\label{x_pdf}
\end{equation}
and the cumulative distribution function (CDF) is
\begin{equation}
	F(x) = 1 - e^{-x/\overline{x}} , \quad x \geq 0.
	\label{x_cdf}
\end{equation}

The selected antenna element of the ST provides the maximum output SINR at the SR. The SR then sends back the index to the ST through an error-free feedback channel. In order to limit the interference caused by the SU, an interference power constraint, denoted by $Q$, is adopted, which limits the instantaneous interference generated by the SU to the PU. In addition, $P_{\max}$ denotes the maximum transmit power of the ST due to, for instance, hardware limitations or regulations. On limited power, now, the transmit power of the ST using the $i$-th antenna can be represented as \cite{Li2012d,Suraweera2010}
\begin{equation}
	P_s^{(i)} = \min\left\{\frac{Q}{h_i}, P_{\max}\right\}.
	\label{tran_power}
\end{equation}
Accordingly, when the $i$-th antenna of the ST is used, the received SINR at the SR is represented as
\begin{equation}
%\begin{split}
	\gamma_i  = \min\left\{\frac{Q g_i}{h_i\left(P_p q+\sigma^2\right)}. \frac{P_{\max}g_i}{P_p q +\sigma^2}\right\} %\\
		%& = \begin{cases}\frac{Q g_i}{h_i\left(P_p q +\sigma^2\right)},& \quad  h_i \geq \frac{Q}{P_{\max}} \\ \frac{P_{\max}g_i}{P_p q+\sigma^2},& \quad  h_i < \frac{Q}{P_{\max}} \end{cases}.
%\end{split}
\label{eq:gamma_i}
\end{equation}
The CDF of $\gamma_i$ has been deduced in \cite[eq. (15)]{Suraweera2010} as
\begin{equation}\begin{split}
	F_{\gamma_i}(x) &= 1 - \frac{Q\overline{g}_i}{x P_p\overline{q} \overline{h}_i} e^{\frac{\sigma^2}{P_p\overline{q}} + \frac{Q\overline{g}_i}{x P_p\overline{q} \overline{h}_i}} \\
		& \quad \times \Gamma\left[0, \left(\frac{x}{\overline{g}_i P_{\max}}+ \frac{1}{P_p\overline{q}}\right)\left(\frac{Q\overline{g}_i}{x \overline{h}_i} + \sigma^2\right)\right] \\
		& \quad - \left[1-e^{-\frac{Q}{\overline{h}_i P_{\max}}}\right] \frac{\overline{g}_i P_{\max}}{x P_p \overline{q} + \overline{g}_i P_{\max}} e^{- \frac{x\sigma^2}{\overline{g}_i P_{\max}}},
	\end{split}
	\label{cdf_gamma_i}
\end{equation}
where $\Gamma(a,b) = \int_b^{\infty} e^{-t} t^{a-1} \mathrm{d} t$ denotes the incomplete Gamma function \cite[eq. 8.350-2]{Gradshteyn2007}. If the TAS technique is used, then the antenna of the ST providing the maximum SINR at the SR is selected such that
\begin{equation}
	\gamma_{\max} = \max_i \gamma_i, \,\, i=1,\cdots,N ,  \label{eq:gamma_max}
\end{equation}
where $\gamma_i$ is given by (\ref{eq:gamma_i}). The CDF of the received SINR at the SR can be written, using order theory \cite{David2003}, as
\begin{equation}
	F_{\gamma_{\max}}(x) = \left[F_{\gamma_i}(x)\right]^{N} .
	\label{cdf_sinr}
\end{equation}
To the best of our knowledge, there is no explicit expressions for mean and outage capacities when using the CDF (\ref{cdf_sinr}) due to its algebraic complexity. In order to obtain tractable analysis and gain insight for the considered cognitive radio system, we deduce asymptotic expressions for the mean capacity and outage capacity using extreme value theory.

\section{Implications of the Extreme Value Theory} \label{sec:EVT}
Let $X_1, \cdots, X_N$ denote the i.i.d. random variables, and let $F_X(x)$ be the underlying cumulative distribution function. We denote by $M_X$ the maximum of $\left\{X_i\right\}_{1\leq i \leq N}$. If there exist real constants $a_N$ and $b_N>0$, and a non-degenerate limit distribution $G(x)$ such that
\begin{equation}
	\lim_{N\rightarrow \infty}\frac{M_X - a_N}{b_N} \xrightarrow[]{d} G(x), % \text{ convergences in distribution to }
	\label{evt_conv}
\end{equation}
then $G(x)$ is one of the extreme value distributions: Gumbel, Fr\'{e}chet, or reverse-Weibull \cite{Haan2006}. We say that $F_X(x)$ belongs to the maximum domain of attraction (MDA) of $G(x)$. 

Before using the extreme value distributions to approximate $F_{\gamma_{\max}}(x)$ in (\ref{cdf_sinr}), we recall a lemma proved in \cite{Song2006, Galambos1978} which gives a sufficient condition for a distribution belonging to the MDA of the Gumbel distribution.

\begin{lemma}\label{lemma_mda} \cite[Theorem 2.7.2]{Galambos1978} and \cite[Lemma 1]{Song2006}. Let $\omega(F_X)=\sup\left\{x: F_X<1\right\}$. Assume that there is a real number $x_1$, such that for $x_1\leq x <\omega(F_X)$, $f_X(x) = F_X'(x) \neq 0$ and $F_X''(x)$ exists. If
	\begin{equation}
		\lim_{x\rightarrow \omega(F_X)}\frac{d}{dx}\left[\frac{1-F_X(x)}{f_X(x)}\right] = 0,
		\label{lemma_evt}
	\end{equation}
then $F_X(x)$ lies on the MDA of the Gumbel distribution (\ref{evt_conv}) with the CDF expressed as
\begin{equation}
	G(x) = e^{-e^{-\frac{x-a_N}{b_N}}},
	\label{eq:CDF_Gumbel}
\end{equation}
where the normalizing constants $a_N$ and $b_N$ are determined by
	\begin{equation}
\begin{split}
	a_N &= F_X^{-1}\left(1-\frac{1}{N}\right), \\
	b_N &= F_X^{-1}\left(1-\frac{1}{Ne}\right) - a_N.
\end{split}
\label{eq:SINR_coeffs}
\end{equation}
Here $e$ is the base of the natural logarithm, and $F_X^{-1}(\cdot)$ denotes the inverse of the CDF. 
\end{lemma}

Using Lemma \ref{lemma_mda}, we prove that the CDF of SINR $F_{\gamma_i}(\cdot)$ lies on the MDA of the Gumbel distribution.

\begin{proposition} \label{proposition_SINR}
	Assume that the transmit power of the secondary transmitter is given by (\ref{tran_power}). Then the SINR distribution $F_{\gamma_i}(\cdot)$ of the SR with transmit antenna selection belongs to the MDA of the Gumbel distribution.
\end{proposition}

\begin{IEEEproof}
see Appendix \ref{append_proof}.
\end{IEEEproof}
Using Proposition \ref{proposition_SINR}, we obtain approximations for the mean capacity and outage capacity of the SU.

\section{Asymptotic Performance Analysis} \label{sec:perf}
%We investigate the asymptotic performance of the SU in presence of Rayleigh fading and interference from the PU. 

%
\subsection{SU Mean Capacity}\label{sec:ec}
%The exact SU mean capacity (nats/s/Hz) can be written as
%\begin{equation}
	%C_{\text{SU}} = \int_0^{\infty}\log\left(1+\gamma\right) \mathsf{d} F_{\gamma_{\max}}(\gamma),
%\end{equation}
%where the CDF $F_{\gamma_{\max}}(\gamma) = \left[F_{\gamma_i}(\gamma)\right]^{N}$ and $F_{\gamma_i}(\gamma)$ is given in (\ref{cdf_gamma_i}).
%
%First, we review the limiting throughput theorem (LTD) proposed in \cite{Song2006}.
%\begin{lemma}\cite{Song2006}
	%Let $\Gamma_i$ denote the received SINR at the secondary receiver when $i^{th}$ is used. Assume that $\Gamma_i's, \quad i=1,\cdots, N$ are i.i.d. with a distribution $F_{\gamma_i}(\gamma)$ and the supremum $\omega(F)\rightarrow \infty$. The distribution of throughput lies on the domain of the attraction of the Gumbel distribution if the distribution of the SINR belongs to the domain of the attraction of the Gumbel distribution.
%\end{lemma}
%

The instantaneous rate for the SU can be represented as
\begin{equation}
	R_{\max} = \max_{1\leq i\leq N}\,\, R_i,  \label{eq:R_max}
\end{equation}
where $R_i=\log\left(1+\gamma_i\right), \forall i=1,\cdots, N$ is the instantaneous rate using the $i$-th antenna. According to the limiting throughput distribution (LTD) theorem proved in \cite{Song2006}, the distribution of $R_i$ belongs to the MDA of the Gumbel distribution. Consequently, we have $\lim_{N\rightarrow \infty}\frac{R_{\max} - a_N}{b_N} \xrightarrow[]{d} G(x)$ and $G(x)$ is expressed in (\ref{eq:CDF_Gumbel}) with the following normalizing constants
\begin{equation}
\begin{split}
	a_N &= \log\left[1+ F^{-1}_{\gamma_i}\left(1-\frac{1}{N}\right)\right], \\
	b_N &= \log\left[\frac{1+ F^{-1}_{\gamma_i}\left(1-\frac{1}{N e}\right)}{1+ F^{-1}_{\gamma_i}\left(1-\frac{1}{N}\right)}\right],
\end{split}
\label{eq:coeff_cap}
\end{equation}
where $F^{-1}_{\gamma_i}(\cdot)$ denotes the inverse function of $F_{\gamma_i}(\cdot)$ in (\ref{cdf_gamma_i}).  
Based on the Lemma 2 in \cite{Song2006}, we find that the expectation of $\frac{R_{\max} - a_N}{b_N}$ converges, and the mean capacity of the SU $C_s = \mathbb{E}\left[R_{\max}\right]$ can be thus approximated by integrating $R_{\max}$ over (\ref{eq:CDF_Gumbel}) as
\begin{equation}
	C_s \approx a_N + b_N\text{E}_0, \label{eq:Cs_appr}
\end{equation}
where $\text{E}_0 = 0.5772\ldots$ denotes the Euler constant. According to Proposition \ref{proposition_SINR} the approximation is tight in the asymptotic regime $N\rightarrow \infty$.

%
%\subsection{SU Mean Capacity Scaling}
Based on (\ref{eq:Cs_appr}), we study the behavior of the SU mean capacity in two limiting regimes. First, we consider that the maximum transmit power constraint approaches infinity ($P_{\max}\rightarrow \infty$). And the second case is that the interference power constraint is much larger than $P_{\max}$, i.e., $Q \gg P_{\max}$. We name these two scenarios as interference power limited regime (IPLR) and transmit power limited regime (TPLR), respectively.

\subsubsection{Interference power limited regime}

When $P_{\max}\rightarrow \infty$ and the $i$-th antenna is selected, we rewrite the asymptotic CDF (\ref{cdf_gamma_i}) of the SINR at the SR as
\begin{equation}
	F_{\gamma_i, \text{IPLR}}(x) = 1-c_p \frac{Q \overline{g}}{x\overline{h}},  % 1-\frac{e^{\sigma^2/\left(P_p \overline{q}\right)} \Gamma\left(0,\frac{\sigma^2}{P_p \overline{q}}\right) \frac{Q}{\overline{h}} \overline{g} }{x P_p \overline{q}}
\end{equation}
where $c_p = e^{\sigma^2/\left(P_p \overline{q}\right)} \Gamma\left(0,\frac{\sigma^2}{P_p \overline{q}}\right)/P_p\, \overline{q}$ represents the effects of the additive noise and the PU interference. The inverse CDF of the SINR is given by
\begin{equation}
	F_{\gamma_i,\text{IPLR}}^{-1}[y] = c_p \frac{Q \overline{g}}{(1-y)\overline{h}}. \label{eq:F_IPLR}
\end{equation}
Using (\ref{eq:coeff_cap}) and (\ref{eq:F_IPLR}), the normalizing constants of the Gumbel distribution $G(x)$ are obtained as
\begin{equation}
	a^{\text{IPLR}}_N = \log\left(1+c_p \frac{Q \overline{g}}{\overline{h}} N\right),	
\label{eq:aN_IPLR}
\end{equation}
and 
\begin{equation}
	b^{\text{IPLR}}_N = \log\left(\frac{1+c_p \frac{Q \overline{g}}{\overline{h}} N e}{1+c_p \frac{Q \overline{g}}{\overline{h}} N}\right),
\label{eq:bN_IPLR}
\end{equation}
where $b^{\text{IPLR}}_N\rightarrow 1$ as $N\rightarrow \infty$. Thus, the asymptotic mean capacity (\ref{eq:Cs_appr}) becomes
\begin{equation}
	C^{\text{IPLR}} = \log\left(1+c_p \frac{Q \overline{g}}{\overline{h}} N\right) +  \text{E}_0, \label{eq:C_IPLR}
\end{equation}
which indicates that the SU mean capacity scales as $\log(N)$ in the interference power limited regime.

%Note that at very low PU interference we have $c_p\approx 1/\sigma^2$, where we use the identity $\lim_{x\rightarrow 0}\frac{e^{1/x}\Gamma(0,1/x)}{x} = 1$. 

%
\subsubsection{Transmit power limited regime}
When $Q \gg P_{\max}$ and the $i$-th antenna is selected, we rewrite the asymptotic CDF (\ref{cdf_gamma_i}) of the SINR at the SR as
\begin{equation}
	F_{\gamma_i, \text{TPLR}}(x) = 1-\frac{c_q\, P_{\max} \overline{g}}{x P_p \overline{q} + P_{\max}\overline{g}} e^{-\frac{x \sigma^2}{P_{\max}\overline{g}}},  \label{eq:F_TPLR}
		%\\
		%& \approx 1-\frac{P_{\max} \overline{g}}{x P_p \overline{q} + P_{\max}\overline{g}} e^{-\frac{x \sigma^2}{P_{\max}\overline{g}}}
\end{equation}
where $c_q = 1-e^{-\rho/\overline{h}}$ and $\rho = Q/P_{\max}$. The inverse CDF of the SINR can be obtained as
\begin{equation}
	F_{\gamma_i, \text{TPLR}}^{-1}[y] \approx \frac{P_{\max}\overline{g}}{\sigma^2} \mathcal{W}\left(\frac{c_q\, e^{\frac{\sigma^2}{P_p \overline{q}}}}{(1-y) P_p \overline{q}/\sigma^2}\right) - \frac{P_{\max}\overline{g}}{P_p \overline{q}},  \label{eq:Finvs_TPLR}
\end{equation}
where $\mathcal{W}(x)$ denotes the Lambert W-Function satisfying $\mathcal{W}(x) e^{\mathcal{W}(x)} = x$. Thus, using (\ref{eq:coeff_cap}) and (\ref{eq:Finvs_TPLR}), the normalizing constants of the Gumbel distribution $G(x)$ are obtained as 
\small
\begin{IEEEeqnarray}{ll}
	a_N^{\text{TPLR}} &= \log\left(1+\frac{P_{\max}\overline{g}}{\sigma^2} \mathcal{W}\left(\frac{N\,c_q\,  e^{\frac{\sigma^2}{P_p \overline{q}}}}{ P_p \overline{q}/\sigma^2}\right) - \frac{P_{\max}\overline{g}}{P_p \overline{q}}\right), \label{eq:aN_TPLR} \\
	b_N^{\text{TPLR}} &= \log\left(1+\frac{P_{\max}\overline{g}}{\sigma^2} \mathcal{W}\left(\frac{e N\,c_q\,  e^{\frac{\sigma^2}{P_p \overline{q}}}}{ P_p \overline{q}/\sigma^2}\right) - \frac{P_{\max}\overline{g}}{P_p \overline{q}}\right)-a_N^{\text{TPLR}}, \nonumber\\ 
	& \label{eq:bN_TPLR}
\end{IEEEeqnarray}
 \normalsize
where $b_N^{\text{TPLR}} = O\left(1/N\right)$ as $N\rightarrow \infty$, and $O\left(z\right)^m$ represents a term of order $\left(z\right)^m$. Therefore, the SU asymptotic mean capacity (\ref{eq:Cs_appr}) becomes
\begin{equation}
	C^{\text{TPLR}} = a_N^{\text{TPLR}} + b_N^{\text{TPLR}} \text{E}_0, \label{eq:C_TPLR}
\end{equation}
which admits an explicit expression and avoids the operation of inverting the SINR CDF function (\ref{cdf_gamma_i}).
% \log\left(1+\frac{P_{\max}\overline{g}}{\sigma^2} \mathcal{W}\left(\frac{N\,c_q\, e^{\frac{\sigma^2}{P_p \overline{q}}}}{ P_p \overline{q}/\sigma^2}\right) - \frac{P_{\max}\overline{g}}{P_p \overline{q}}\right)

%Since the LambertW function $\mathcal{W}(x)\approx \log(x)$ as $x\rightarrow \infty$ ??????????????????????????????????????????, which indicates that the SU mean capacity scales as $\log\log(N)$ in the transmit power limited regime. 

When the interference from the PU can be ignored, i.e. $P_p \overline{q}\rightarrow 0$, (\ref{eq:F_TPLR}) becomes
\begin{equation}
	F_{\gamma_i, \text{TPLR}}(x) \approx 1- c_q e^{-\frac{x \sigma^2}{P_{\max}\overline{g}}},   % Expanding at p->0
\end{equation}
which is the approximation of the CDF of the SINR at low interference regime. Then we obtain the inverse CDF as
\begin{equation}
	F_{\gamma_i, \text{TPLR}}^{-1}[y] \approx \frac{P_{\max}\overline{g}}{\sigma^2}\log\left(\frac{c_q}{1-y}\right).
	\label{eq:TPLR}
\end{equation}
Substituting (\ref{eq:TPLR}) into (\ref{eq:coeff_cap}), we have the normalizing constants 
\begin{equation}
	a_{N, \text{low}}^{\text{TPLR}} = \log\left(1+\frac{P_{\max}\overline{g}}{\sigma^2}\log(c_q N)\right), \label{eq:aN_TPLR_low}
\end{equation} 
\begin{equation}
	b_{N, \text{low}}^{\text{TPLR}} = \log\left(1+\frac{P_{\max}\overline{g}/\sigma^2}{1+\frac{P_{\max}\overline{g}}{\sigma^2}\log(c_q N)}\right). \label{eq:bN_TPLR_low}
\end{equation} 
Substituting (\ref{eq:aN_TPLR_low}) and (\ref{eq:bN_TPLR_low}) into (\ref{eq:Cs_appr}), the asymptotic mean capacity is obtained as
\begin{equation}
	C^{\text{TPLR}}_{\text{low}} \approx a_{N, \text{low}}^{\text{TPLR}} + b_{N, \text{low}}^{\text{TPLR}} \text{E}_0,	 \label{eq:C_TPLR_noPp}
\end{equation}
where $b_{N, \text{low}}^{\text{TPLR}} = O\left(1/N\right)$ as $N\rightarrow \infty$. This shows that the SU mean capacity scales as $\log\log(N)$ in the transmit power limited regime with low PU interference. Note that in the TPLR with low PU interference, the considered model (\ref{eq:signal_model}) reduces to the conventional point-to-point communication. In this case, the derived result (\ref{eq:C_TPLR_noPp}) is in line with \cite[eq. (38)]{Bai2009}.

\subsubsection*{Remarks} %The asymptotic received signal to noise ratio can be written as
%\begin{equation}
	%\text{SNR} = \frac{\min\left[\log(N)P_{\max}, N Q\right]}{\sigma^2}\overline{g}.
%\end{equation}
%Such that the asymptotic SU mean capacity at very low interference becomes
%\begin{equation}
	%\text{C}_{\text{asympt}} = \log\left(1+\frac{\min\left[\log(N)P_{\max}, N Q\right]}{\sigma^2}\overline{g}\right) \label{eq:C_asympt}
%\end{equation}
The results suggest that the TAS provides different power gains in these two scenarios. In the TPLR, (\ref{eq:C_TPLR_noPp}) shows that increasing the number of transmit antennas may not be an efficient way to achieve higher mean capacity, while it is effective in the IPLR as shown in (\ref{eq:C_IPLR}).

\subsection{SU Outage Capacity}\label{sec:outc}
Given the rate $r$ (nats/s/Hz), the SU outage probability after transmit antenna selection can be defined as 
\begin{equation}
	F_{\text{out}}(r) = \Pr\left\{R_{\max}<r\right\},
\end{equation}
where $R_{\max}$ is given by (\ref{eq:R_max}). The outage capacity can be interpreted as the maximum rate such that $F_{\text{out}}\leq \epsilon$, where $\epsilon$ is the defined outage threshold. For monotonically increasing function $F_{\text{out}}(r)$, we have
\begin{equation}
	C_{\text{out}} = F_{\text{out}}^{-1}(\epsilon). \label{eq:cout_def} %  \max\left\{r: \,\,\Pr\left\{R_{\max}<r\right\} < \epsilon\right\}
\end{equation}
%Obtaining the above exact outage capacity requires integrating with respect to $\gamma_{\max}$, which is algebraically complex. In order to have tractable analysis and gain insight for the considered cognitive radio system, we obtain the asymptotic expressions for the outage capacity.
According to the LTD theorem \cite{Song2006}, the asymptotic distribution of $R_{\max}$ can be approximated as (\ref{eq:CDF_Gumbel}). Thereafter, we have the following proposition for the SU outage capacity.
\begin{proposition} \label{proposition_C_out}
	With TAS, the asymptotic outage capacity of the SU is given by
	\begin{equation}
	C_{\text{out}} \approx a_N - b_N \log\log\left(\frac{1}{\epsilon}\right), \label{eq:C_out}
\end{equation}
where the normalizing constants $a_N$ and $b_N$ are the normalizing coefficients given in (\ref{eq:coeff_cap}).
\end{proposition}

\begin{IEEEproof}
	The proof of this proposition is straightforward by substituting (\ref{eq:CDF_Gumbel}) into (\ref{eq:cout_def}).
\end{IEEEproof}
Based on Proposition \ref{proposition_C_out}, we investigate the SU outage capacity scaling properties in case of IPLR and TPLR, respectively.

\subsubsection{Interference power limited regime} When $P_{\max}\rightarrow \infty$, the normalizing constants $a_N^{\text{IPLR}}$ and $b_N^{\text{IPLR}}$ are computed in (\ref{eq:aN_IPLR}) and (\ref{eq:bN_IPLR}), respectively.  Following (\ref{eq:C_out}), the asymptotic SU outage capacity in IPLR yields

\small
\begin{equation}
	C^{\text{IPLR}}_{\text{out}} = \log\left(1+c_p \frac{Q \overline{g}}{\overline{h}} N\right) -  \log\left(\frac{1+c_p \frac{Q \overline{g}}{\overline{h}} N e}{1+c_p \frac{Q \overline{g}}{\overline{h}} N}\right)\log\log\left(\frac{1}{\epsilon}\right). \label{C_out_IPLR}
\end{equation} 

\normalsize
Recall that $b_N^{\text{IPLR}} \rightarrow 1$ when $N\rightarrow \infty$, we have
\begin{IEEEeqnarray}{ll}
	C^{\text{IPLR}}_{\text{out}} &\approx \log\left(1+c_p \frac{Q \overline{g}}{\overline{h}} N\right) - \log\log\left(\frac{1}{\epsilon}\right) \\
		& =  C^{\text{IPLR}} - \text{E}_0 - \log\log\left(\frac{1}{\epsilon}\right), \label{eq:C_IPLR_out}
\end{IEEEeqnarray}
where $C^{\text{IPLR}}$ is given in (\ref{eq:C_IPLR}). We can see that the SU outage capacity scales in the same manner as the SU mean capacity. When given the outage threshold $\epsilon$, there exists a constant gap $\text{E}_0 + \log\log\left(\frac{1}{\epsilon}\right)$ between $C^{\text{IPLR}}_{\text{out}}$ and $C^{\text{IPLR}}$. % The SU outage capacity scales with $\log(N)$.

\subsubsection{Transmit power limited regime}
When $Q\gg P_{\max}$, following (\ref{eq:C_out}) the asymptotic SU outage capacity in TPLR yields
\begin{equation}
C^{\text{TPLR}}_{\text{out}} = a_N^{\text{TPLR}} - b_N^{\text{TPLR}}\log\log\left(\frac{1}{\epsilon}\right), \label{eq:c_out_TPLR}
\end{equation}
where the normalizing constants $a_N^{\text{TPLR}}$ and $b_N^{\text{TPLR}}$ are computed in (\ref{eq:aN_TPLR}) and (\ref{eq:bN_TPLR}), respectively. 
We can see that the SU outage capacity scales the same way as the SU mean capacity, i.e. $C^{\text{TPLR}}_{\text{out}}\sim \log\log\left(N\right)$ as $N\rightarrow \infty$. It is also shown that the asymptotic SU outage capacity does not depend on the outage threshold $\epsilon$. However, the convergence speed is pretty slow due to $C^{\text{TPLR}}_{\text{out}}\sim \log\log\left(N\right)$.

When the interference from the PU can be ignored in TPLR, i.e. $P_p \overline{q}\rightarrow 0$, the asymptotic SU outage capacity (\ref{eq:c_out_TPLR}) yields
\begin{equation}
	C^{\text{TPLR}}_{\text{out, low}} \approx a_{N, \text{low}}^{\text{TPLR}} - b_{N, \text{low}}^{\text{TPLR}} \log\log\left(\frac{1}{\epsilon}\right). \label{eq:C_out_TPLR_low}
\end{equation}
where $a_{N, \text{low}}^{\text{TPLR}}$ and $b_{N, \text{low}}^{\text{TPLR}}$ are given by (\ref{eq:aN_TPLR_low}) and (\ref{eq:bN_TPLR_low}), respectively.

\subsubsection*{Remarks} The number of antennas has different influence on the asymptotic outage capacity of the SU. In the TPLR case, i.e. $P_{\max}\ll Q$,  the SU outage capacity increases logarithmically with $N$ and the gap between the SU mean capacity and outage capacity vanishes.  However, this is not for the IPLR scenario, i.e. $P_{\max}\rightarrow \infty$, that the outage capacity has a much slower scaling rate, $\log\log(N)$. In addition, there is always a gap between the outage capacity $C^{\text{IPLR}}_{\text{out}}$ and mean capacity $C^{\text{IPLR}}$.

% %
\section{Simulation Results}\label{sec:Sim}
We present the asymptotic results with respect to $N$ for the SU mean capacity and the outage capacity, assuming that the ST employs transmit antenna selection and the transmit power is restricted by the peak interference power constraint and the maximum transmit power constraint. The rate scaling behaviors of the mean capacity and outage capacity are also discussed. The mean channel power gains are set to 1. The noise power per Hz at the SR is set to $-10$ dB, i.e., $\sigma^2 = 0.1$, where we assume that the maximum average received signal to noise ratio at the SR is greater or equal to 1, i.e., $\mathbb{E}\left[P_{\max}\overline{g_i}\right]/\sigma^2\geq 1$. Let $\mathsf{INR}=P_p/\sigma^2$ denote the interference to the noise ratio, $\mathsf{SIR_Q}=Q/P_p$ and $\mathsf{SIR_P}=P_{\max}/P_p$ be the transmit signal to the interference power ratio, and $\mathsf{SNR_Q}=Q/\sigma^2$ and $\mathsf{SNR_P}=P_{\max}/\sigma^2$ be the transmit signal to the noise power ratio. In addition, $\mathsf{PQR}=P_{\max}/Q$ is the maximum power to the interference constraint ratio.

\begin{figure}[!t]
	\centering
		\includegraphics[width=0.7\columnwidth]{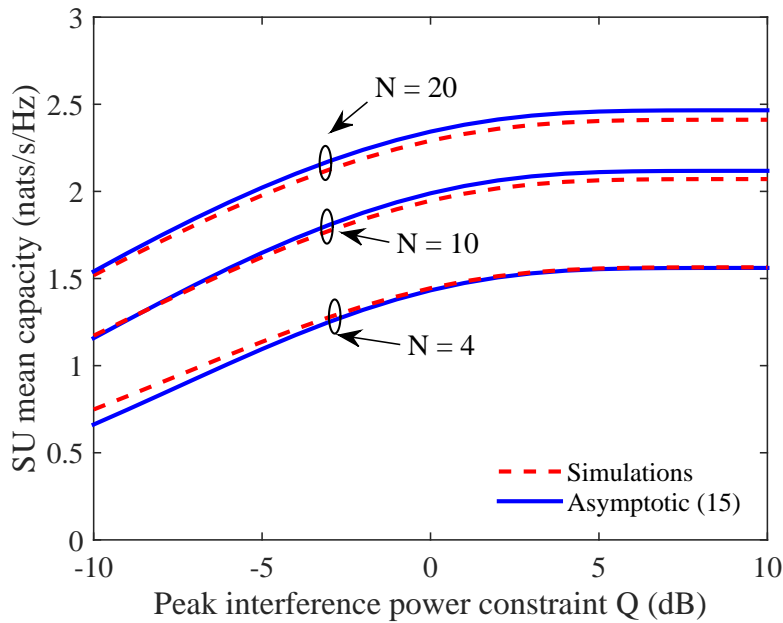}
	\caption{SU mean capacity versus the peak interference power constraint $Q$ with $P_{\max} = 0$ dB and $P_p = 0$ dB.}
	\label{fig:ergC_Q}
\end{figure}

Fig. \ref{fig:ergC_Q} shows the SU mean capacity versus the interference power constraint $Q$ when the number of the ST antennas is $N=4$, 10, and 20. The PU transmit power is set to $P_p=0$ dB, and the maximum transmit power $P_{\max}$ of the SU is $0$ dB. We also plotted the asymptotic mean capacity of the SU according to (\ref{eq:Cs_appr}), and compared it to the simulation results generated by $10^{6}$ channel realizations. As the interference power constraint $Q$ increases, the SU mean capacity is improved until the maximum transmit power constraint $P_{\max}$ becomes dominating. We can see that the asymptotic results using EVT are reasonably accurate, which validates our approach to characterize the SU mean capacity using simple and explicit expressions.

\begin{figure}[!t]
	\centering
		\subfloat[IPLR]{\label{fig:c_Scal_IPLR} \includegraphics[width=0.7\columnwidth]{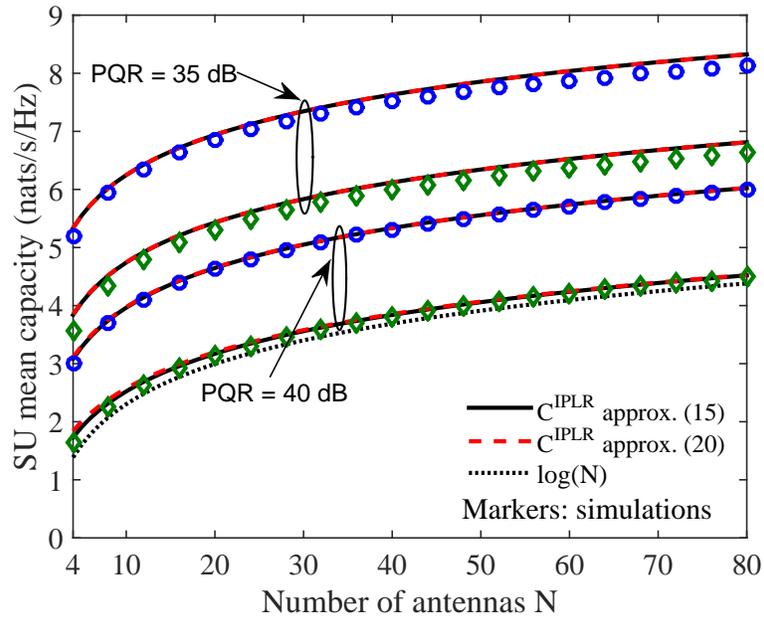}} \\
		\subfloat[TPLR]{\label{fig:c_Scal_TPLR} \includegraphics[width=0.7\columnwidth]{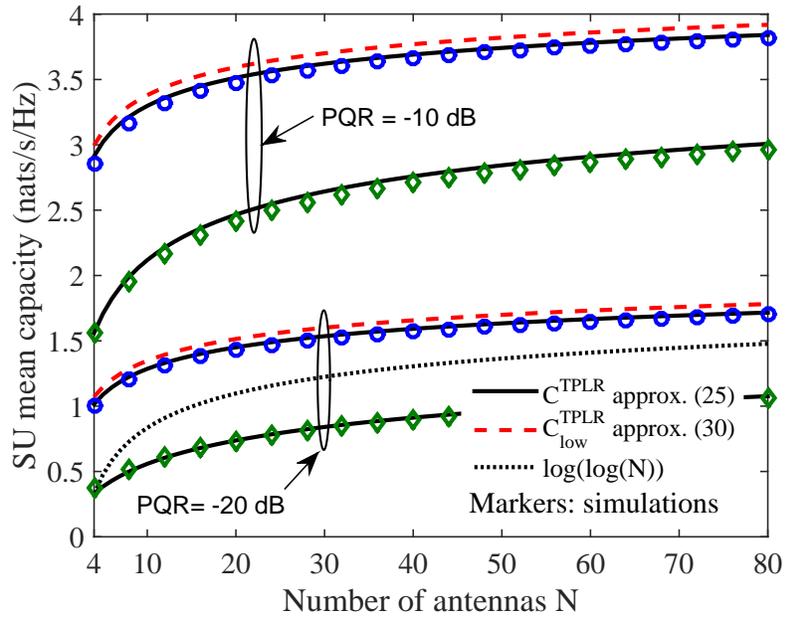}} 
	\caption{SU mean capacity versus the number of ST antennas $N$. Lines with circles denote $\mathsf{INR} = -10$ dB; lines with diamonds denote  $\mathsf{INR} = 10$ dB. (a) IPLR: $\mathsf{PQR} = \left\{35, 40\right\}$ dB.  (b) TPLR: $\mathsf{PQR} = \left\{-10, -20\right\}$ dB.}
	\label{fig:c_Scal}
\end{figure}

Fig. \ref{fig:c_Scal} shows the the SU mean capacity as a function of the number of ST antennas in the interference power and the transmit power limited regimes, respectively. In Fig. \ref{fig:c_Scal_IPLR}, the maximum transmit power constraint is set to $P_{\max}=30$ dB$\gg Q$. In this case, we calculate the SU mean capacity using IPLR approximation, namely (\ref{eq:Cs_appr}) with the normalizing constant $a^{\text{IPLR}}_N$ and $b^{\text{IPLR}}_N$ given by (\ref{eq:aN_IPLR}) and (\ref{eq:bN_IPLR}), respectively. In addition, we show the SU rate scaling (\ref{eq:C_IPLR}) in IPLR. Numerical results show that the approximation and scaling yield good agreement with simulations, especially when $\mathsf{PQR}=35$ dB and $\mathsf{SNR_Q} = 5$ dB. Furthermore, the scaling behavior of the SU capacity is shown to be $\log(N)$ as predicted by (\ref{eq:C_IPLR}). In Fig. \ref{fig:c_Scal_TPLR}, we assume that $\mathsf{PQR} = \left\{-10, -20\right\}$ dB. The SU mean capacity is computed using TPLR approximation (\ref{eq:C_TPLR}), and the SU rate scaling is obtained via (\ref{eq:C_TPLR_noPp}). In TPLR with low PU interference, e.g. $\mathsf{INR} = -10$ dB, the SU capacity scales as $\log(\log(N))$. Results show that in TPLR, it is more effective to improve the SU capacity by increasing the transmit power than by using more transmit antennas. However, in IPLR if the SU is subject to a lower interference limit, the mean capacity can be maintained by using more transmit antennas.

\begin{figure}[!t]
	\centering
		\includegraphics[width=0.7\columnwidth]{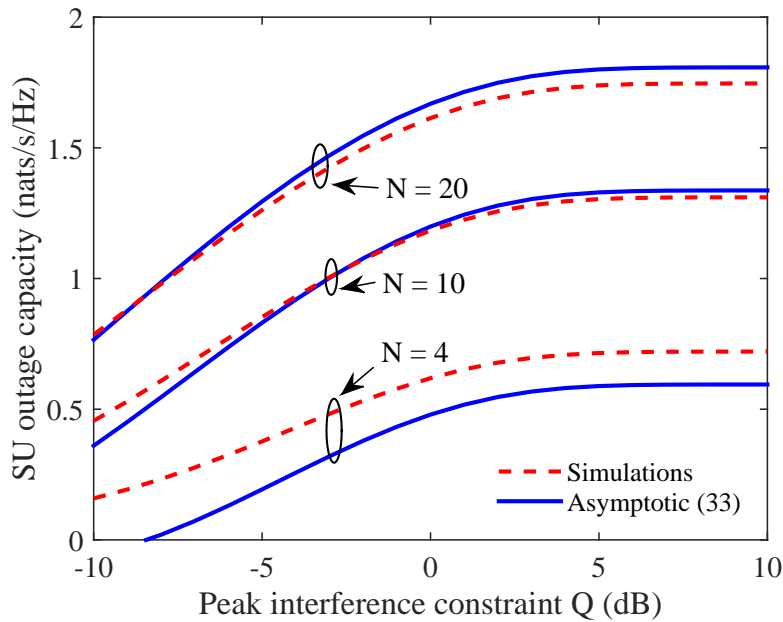}
	\caption{SU outage capacity. $P_{\max}=0 dB$, $P_p=0dB$, and $\epsilon = 10\%$.}
	\label{fig:c_out_Q}
\end{figure}

The SU outage capacity is plotted as a function of the interference constraint $Q$  in Fig. \ref{fig:c_out_Q}, where the outage probability is set to $10\%$. Although the EVT approximation is not very accurate for $N=4$ ST antennas, it gives an estimate trend of the SU outage capacity. As the number of antennas increases, the accuracy of the EVT approximation improves. In addition, for $N=20$ the approximation for a large range interference constraint, $Q\geq -10$dB, is acceptably accurate for understanding such systems.
  
\begin{figure}[!t]
	\centering
		\subfloat[IPLR]{\label{fig:out_IPLR}\includegraphics[width=0.7\columnwidth]{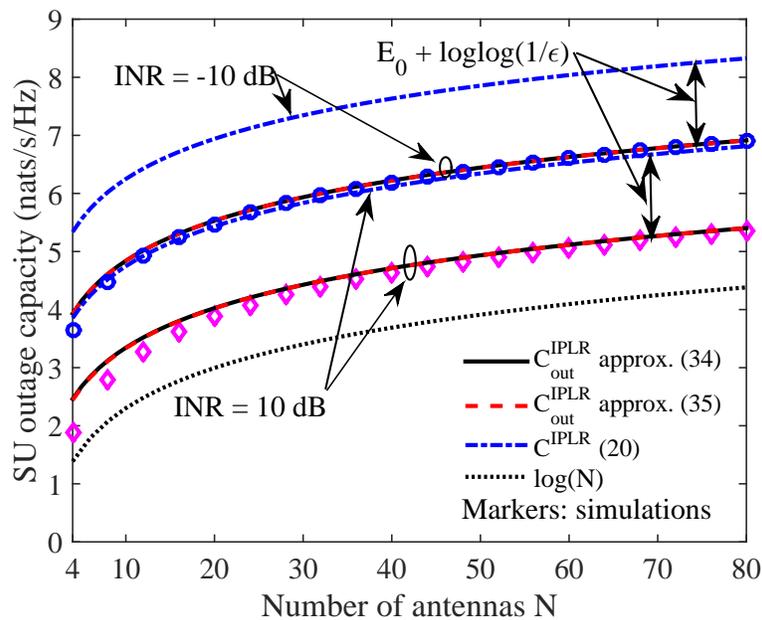}}\\
		\subfloat[TPLR]{\label{fig:out_TPLR}\includegraphics[width=0.7\columnwidth]{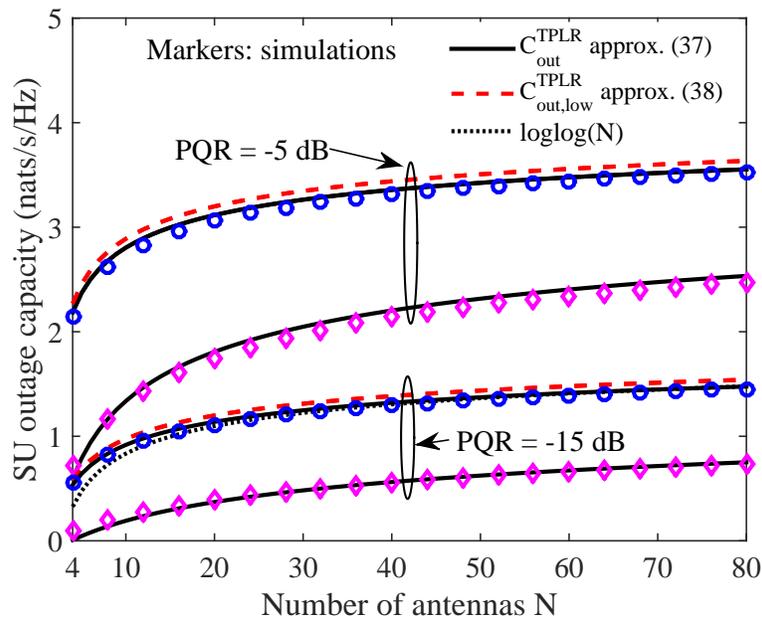}}
	\caption{SU outage capacity versus the number of ST antennas $N$. Lines with circles denote $\mathsf{INR}=-10$ dB; lines with diamonds denote $\mathsf{INR}=10$ dB. The peak interference power constraint $Q = 5$ dB. Outage threshold $\epsilon = 10\%$. (a) IPLR: $\mathsf{PQR}=25$ dB. (b) TPLR: $\mathsf{PQR} = \left\{-5, -15\right\}$ dB.}
	\label{fig:c_outN_scal}
\end{figure}

In order to gain some insight into the asymptotic behavior of the SU outage capacity, we depict in Fig. \ref{fig:c_outN_scal} the SU outage capacity in the transmit power limited and interference power limited regimes, respectively. The asymptotic SU mean capacity is also plotted for a comparison. As the number of ST  antennas increases, the asymptotic outage capacities for various cases are accurate compared to the numerical simulations. The scaling properties of the SU outage capacity are validated. In Fig. \ref{fig:out_IPLR}, the SU outage capacity in IPLR behaves as $\log(N)$ when the number of ST antennas $N$ approaches to infinity. In this case, there is a constant gap $\text{E}_0 +\log\log\left(1/\epsilon\right)$ between the outage and the mean capacities as predicted by (\ref{eq:C_IPLR_out}). Yet, the SU outage capacity in TPLR behaves as $\log\log(N)$ as shown in Fig. \ref{fig:out_TPLR}, which can be proved using (\ref{eq:C_out_TPLR_low}).

% %
\section{Conclusion}\label{sec:Con}
In this paper, we applied the extreme value theory while studying the performance of a spectrum sharing system in terms of the SU mean capacity, the SU outage capacity, and the rate scaling characteristics. The SU mean capacity and outage capacity have two scaling behaviors with the number of available transmit antennas: in the interference power limited regime (IPLR), capacities scale as $\log(N)$, while they scale as $\log(\log(N))$ in transmit power limited regime (TPLR). These results indicate that the transmit antenna selection technique provides different power gain in considered two scenarios. The obtained accurate approximations provide us a better understanding on the performance of the spectrum-sharing systems with multiple transmit antennas.

% if have a single appendix:
%\appendix[Proof of the Zonklar Equations]
% or
%\appendix  % for no appendix heading
% do not use \section anymore after \appendix, only \section*
% is possibly needed

% use appendices with more than one appendix
% then use \section to start each appendix
% you must declare a \section before using any
% \subsection or using \label (\appendices by itself
% starts a section numbered zero.)
%

\appendix[Derivation of Proposition \ref{proposition_SINR}]
\label{append_proof}

The associated PDF becomes, through taking the derivation to $F_{\gamma_i}(x)$ with respect to $x$ and expanding at $x\rightarrow \infty$,
\begin{equation}
	f_{\gamma_i}(x) = e^{-\frac{\sigma^2 x}{P_{\max}\overline{g}}} \left(\frac{\sigma^2\left(1 - e^{-\frac{Q}{P_{\max}\overline{h}}}\right)}{P_p\overline{q}x} + O\left(\frac{1}{x}\right)^2\right) 
	\label{pdf_gamma_i}
\end{equation}
where $O\left(z\right)^m$ represents a term of order $\left(z\right)^m$. The expansion at $x\rightarrow \infty$ of the secondary derivative, $f'_{\gamma_i}(x)$, can be expressed as
\begin{equation}
	f'_{\gamma_i}(x) = e^{-\frac{\sigma^2 x}{P_{\max}\overline{g}}} \left(-\frac{\sigma^4\left(1 - e^{-\frac{Q}{P_{\max}\overline{h}}}\right)}{P_{\max}\overline{g} P_p\overline{q} x} + O\left(\frac{1}{x}\right)^2\right) 
	\label{pdf_gamma_i_deriv}
\end{equation}

It it obvious, from (\ref{pdf_gamma_i}) and (\ref{pdf_gamma_i_deriv}), that $\lim_{x\rightarrow \infty}f_{\gamma_i}(x) = 0$ and $\lim_{x\rightarrow \infty}\left[-f'_{\gamma_i}(x)\right] = 0$, respectively. We need to show, after taking the derivative in (\ref{lemma_evt}), that
\begin{equation}
	\lim_{x\rightarrow \infty} \left[\frac{(1-F_{\gamma_i}(x))f_{\gamma_i}'(x)}{f_{\gamma_i}^2(x)}\right] = -1
\end{equation}

$F_{\gamma_i}(x)$ is the PDF of $\gamma_i$, thus we have 
\begin{equation}
	\lim_{x\rightarrow \infty}\left[F_{\gamma_i}(x)-1\right] = 0
\end{equation}

Also, we have $\lim_{x\rightarrow \infty}\left[f_{\gamma_i}(x)\right] = 0$. Therefore we obtain, according to L'Hospital's rule, 
\begin{equation}
\begin{split}
	& \lim_{x\rightarrow \infty}\left[\frac{1-F_{\gamma_i}(x)}{f_{\gamma_i}(x)}\right] = \lim_{x\rightarrow \infty}\left[\frac{\left(1-F_{\gamma_i}(x)\right)'}{f_{\gamma_i}'(x)}\right] \\
	& = \lim_{x\rightarrow \infty}\left[\frac{f_{\gamma_i}(x)}{-f_{\gamma_i}'(x)}\right] = \lim_{x\rightarrow \infty}\left(\frac{P_{\max}\overline{g}}{\sigma^2} + O\left(\frac{1}{x}\right)\right) = \frac{P_{\max}\overline{g}}{\sigma^2}
\end{split}
\end{equation}
Thus, $\lim_{x\rightarrow \infty}\left[\frac{1-F_{\gamma_i}(x)}{f_{\gamma_i}(x)}\right]$ or $\lim_{x\rightarrow \infty}\left[\frac{f_{\gamma_i}(x)}{-f_{\gamma_i}'(x)}\right]$ both have a finite limit. Therefore, we have
\begin{IEEEeqnarray}{rl}
	& \lim_{x\rightarrow \infty} \left[\frac{(1-F_{\gamma_i}(x))f_{\gamma_i}'(x)}{f_{\gamma_i}^2(x)}\right] \nonumber\\
	 &\quad = \lim_{x\rightarrow \infty}\left(\frac{1-F_{\gamma_i}(x)}{f_{\gamma_i}(x)}\right)  \lim_{x\rightarrow \infty}\left(\frac{f_{\gamma_i}'(x)}{f_{\gamma_i}(x)}\right) = -1
\end{IEEEeqnarray}
This completes the proof.

%\section{Derivation of Proposition \ref{proposition_C_out}}
%\label{append_proof_p2}
%Using (\ref{eq:CDF_Gumbel}), the maximum rate that satisfies (\ref{eq:cout_def}) with equality can be obtained as
%\begin{equation*}
	%
%\end{equation*}

%% use section* for acknowledgement
%\section*{Acknowledgment}
%
%
%The authors would like to thank...

% Can use something like this to put references on a page
% by themselves when using endfloat and the captionsoff option.
\ifCLASSOPTIONcaptionsoff
  \newpage
\fi

% trigger a \newpage just before the given reference
% number - used to balance the columns on the last page
% adjust value as needed - may need to be readjusted if
% the document is modified later
%\IEEEtriggeratref{8}
% The "triggered" command can be changed if desired:
%\IEEEtriggercmd{\enlargethispage{-5in}}

% references section

% can use a bibliography generated by BibTeX as a .bbl file
% BibTeX documentation can be easily obtained at:
% http://www.ctan.org/tex-archive/biblio/bibtex/contrib/doc/
% The IEEEtran BibTeX style support page is at:
% http://www.michaelshell.org/tex/ieeetran/bibtex/
%\bibliographystyle{IEEEtran}
% argument is your BibTeX string definitions and bibliography database(s)
%\bibliography{IEEEabrv,../bib/paper}
%
% <OR> manually copy in the resultant .bbl file
% set second argument of \begin to the number of references
% (used to reserve space for the reference number labels box)
\bibliographystyle{IEEEtran}
%\bibliography{../../MyResearch}

% biography section
% 
% If you have an EPS/PDF photo (graphicx package needed) extra braces are
% needed around the contents of the optional argument to biography to prevent
% the LaTeX parser from getting confused when it sees the complicated
% \includegraphics command within an optional argument. (You could create
% your own custom macro containing the \includegraphics command to make things
% simpler here.)
%\begin{biography}[{\includegraphics[width=1in,height=1.25in,clip,keepaspectratio]{mshell}}]{Michael Shell}
% or if you just want to reserve a space for a photo:
%
%\begin{IEEEbiography}{Michael Shell}
%Biography text here.
%\end{IEEEbiography}
%
%% if you will not have a photo at all:
%\begin{IEEEbiographynophoto}{John Doe}
%Biography text here.
%\end{IEEEbiographynophoto}
%
%% insert where needed to balance the two columns on the last page with
%% biographies
%%\newpage
%
%\begin{IEEEbiographynophoto}{Jane Doe}
%Biography text here.
%\end{IEEEbiographynophoto}

% You can push biographies down or up by placing
% a \vfill before or after them. The appropriate
% use of \vfill depends on what kind of text is
% on the last page and whether or not the columns
% are being equalized.

%\vfill

% Can be used to pull up biographies so that the bottom of the last one
% is flush with the other column.
%\enlargethispage{-5in}

% that's all folks
\end{document}